\documentclass[aps,prl,twocolumn,amsmath,amssymb,floatfix, superscriptaddress]{revtex4-2}
\usepackage{tabularx}
\usepackage{bm}
\usepackage{euscript}
\usepackage{epsfig,psfrag,subfigure}
\usepackage{graphicx}
\usepackage{color}
\usepackage{amsfonts}
\usepackage{exscale}
\usepackage{amsbsy}
\usepackage{multirow}
\usepackage[normalem]{ulem}
\usepackage[colorlinks=true,linkcolor=blue,citecolor=blue, urlcolor=blue,anchorcolor=blue]{hyperref}

\begin{document}

\title{Pair density wave, infinite-length stripes, and holon Wigner crystal in single-band Hubbard model on diagonal square lattice}
\author{Zhi Xu}
\thanks{These authors contributed equally.}
\affiliation{School of Physical Science and Technology, ShanghaiTech University, Shanghai 201210, China}

\author{Gui-Xin Liu}
\thanks{These authors contributed equally.}
\affiliation{School of Physical Science and Technology, ShanghaiTech University, Shanghai 201210, China}

\author{Yi-Fan Jiang}
\email{jiangyf2@shanghaitech.edu.cn}
\affiliation{School of Physical Science and Technology, ShanghaiTech University, Shanghai 201210, China}

\begin{abstract}
We employ large-scale density-matrix renormalization group (DMRG) simulations to investigate the quantum phase diagram of the hole-doped Hubbard model on square lattices.
By implementing a diagonally oriented square lattice and GPU-accelerated DMRG with up to $48000$ states, we identify three distinct quantum phases across $\delta = 5\%$ to $15\%$ doping: (i) A diagonal stripe phase with short-range uniform superconductivity (SC) at lower doping $\delta\lesssim 9\%$; 
(ii) An intermediate holon Wigner crystal (WC*) phase exhibiting bidirectional charge-density order and short-range SC with spatial oscillating correlations;
(iii) An unprecedented infinite-length stripe (i-stripe) phase at $\delta\gtrsim 12\%$ hosting long stripes spanning the whole lattice. Remarkably, as doping increases, the short-range SC in WC* phase evolves into a 2D-like pair density wave (PDW) with divergent susceptibility in the i-stripe phase, constituting probably the first controlled numerical evidence of dominant PDW in the single-band square-lattice Hubbard model. The established 2D-like PDW and its interplay with charge orders provide new perspectives on dynamical layer decoupling phenomena in cuprates and multifaceted relationships between charge, spin and SC orders in quantum materials.
\end{abstract}

\maketitle

The intricate interplay between spin and charge degrees of freedom in strongly correlated systems leads to a plethora of competing orders including charge density wave (CDW), spin density wave (SDW), and unconventional superconductivity (SC). Of particular interest is the pair-density wave SC---a long-sought superconducting state characterized by finite-momentum Cooper pairs \cite{Agterberg2020}. Distinct from the Fulde-Ferrell-Larkin-Ovchinnikov state in weakly interacting systems under external magnetic fields \cite{Larkin1965, Fulde1964}, PDW could intrinsically emerge from strong electron correlations, without the need to apply external field. 
Recently, PDW has been experimentally observed in a variety of systems such as cuprates \cite{Li2007,Berg2007, Agterberg2008, Berg2009np,Hamidian2016, Ruan2018, Edkins2019, Du2020, Li2021}, iron-based superconductors \cite{Zhao2023,Liu2023}, and other quantum materials \cite{Chen2021,Gu2023,Aishwarya2023,Devarakonda2024}. Its physics is deeply intertwined with charge and spin orders. Understanding the nature of PDW and its relationship with competing orders offers valuable insight into fundamental mechanisms underlying strongly correlated systems \cite{Himeda2002, Wu2007, Raczkowski2007, Aperis2008, Yang2009, Loder2011, Cho2012, You2012, Lee2014, Soto2014, Jian2015, Wang2015, Wardh2017, Han2020, Tommy2020,chakraborty2021odd, Dash2021, wang2024, setty2022, Jin2022, Han2022, Coleman2022, Shaffer2023,  Wu2023Raghu,  Castro2023, Wu2023Yao, JiangBarlas2023, Setty2023, schwemmer2023pair, Fangze2024, Jiang2024pdw, Jiucai2024, Xingchuan2024}. 
Despite its pivotal role, conclusive evidence of PDW in simple strongly correlated models remains largely elusive.  

The Hubbard model, a paradigmatic model for cuprates and correlated electrons, provides fertile ground for exploring competing orders including various forms of antiferromagnetism (AFM), CDW, and unconventional SC \cite{Dagotto1994,Lee2006,Scalapino2012, Fradkin2015,Keimer2015,Arovas2022,Qin2022}. 
Recent studies employing controlled numerical approaches led to a consensus that insulating stripe state \cite{Zaanen1989,Zaanen1998,Machida1989,Kato1990} emerges in lightly doped Hubbard models with only nearest-neighbor (NN) hopping \cite{White1999,Scalapino2012W,Zheng2017,Qin2020,Dodaro2017,Jiang2020Hub,Jiang2024}. Nonetheless, $d$-wave SC and other forms of charge stripes remain energetically competitive with filled stripes \cite{Corboz2011,Corboz2014,Dong2020}. Introducing next-nearest-neighbor (NNN) hopping $t'$ further tips the delicate balance between SC and CDW orders \cite{Dodaro2017, Jiang2018tJ,Jiang2019Hub, Jiang2020Hub,Peng2022, Gong2021, Jiang2021White, Jiang2023,Lu2023Sign, Feng2023,Jiang2024,Lu2024, Xu2024,Yang2024}.
For instance, density matrix renormalization group (DMRG) studies on six-leg cylinders demonstrated that positive $t'$ stabilizes partially filled stripes with quasi-long-range SC \cite{Jiang2020Hub,Jiang2024}, while large negative $t'$ favors a Wigner crystal of holons (WC*) with short-range SC at low doping \cite{Jiang2024}. Constraint-path quantum Monte Carlo study on broader systems suggests uniform SC in both electron- and hole-doped regimes \cite{Xu2024}.
PDW order has recently been observed in related systems like the Emery model on two-leg ladders \cite{Jiang2023threeband,HCJiang2023}. However, obtaining unbiased numerical evidence for dominant SC order in extended Hubbard models on wide lattices and understanding its interplay with other orders remains outstanding challenges.

\begin{figure*}[t]
\centering
\includegraphics[width=\linewidth]{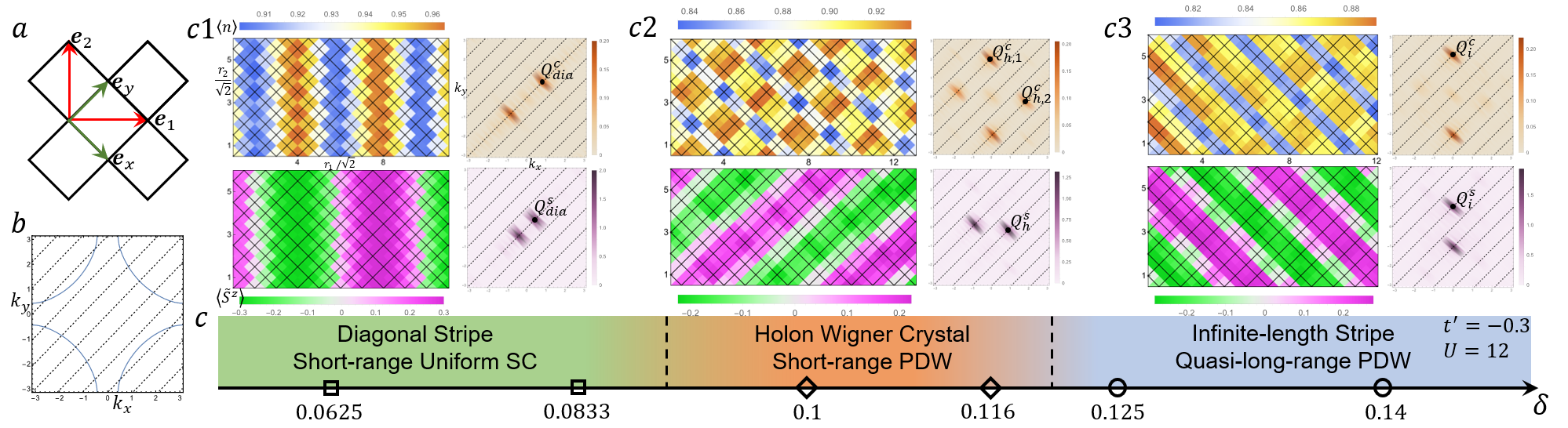}
\caption{(Color online) (a) Illustration of diagonal square lattice. (b) Fermi surface of the $U=0$, $t'=-0.3$ and $\delta=14\%$ model. The dashed lines indicate quantized momenta of $L_2=6$ lattice. (c) Quantum phase diagram of the $U=12$ model with $\delta=5\%\sim 15\%$, including (c1) a diagonal stripe phase (c2) a WC* phase and (c3) an infinite-length stripe (i-stripe) phase with quasi-long-range PDW. These phases are determined by the charge density profile $\left< n(r_1,r_2)\right>$ and spin density profile $ (-1)^{\sqrt{2}r_1}\left< S_z(r_1,r_2) \right>$ and their Fourier transform $n(\bold{k}) =\frac{1}{N} \sum_i (n_i - \bar{n})e^{i \bold{k} \cdot r_i}$ with $\bar{n}=1-\delta$ and $S(\bold{k})=\frac{1}{N} \sum_i  \tilde{S}_i e^{i \bold{k} \cdot r_i}$.}
\label{Fig:phase_diagram}
\end{figure*}

In this paper, we address these challenges by reconfiguring the Hubbard model on a diagonal square lattice \cite{Berg2010diag,Jiang2017} rotated by $\pi/4$ relative to conventional orientations (Fig.~\ref{Fig:phase_diagram}(a)). In finite-size systems, this geometry provides three decisive advantages for probing two-dimensional (2D) physics. First, its unique symmetries allow unambiguous discrimination between $d$-wave and $s$-wave SC order, and between vertical stripes and bidirectional CDW. Second, its quantized momenta intrinsically incorporate a $d$-wave nodal line and more Fermi points as shown in Fig.~\ref{Fig:phase_diagram}(b), enabling low-energy physics exploration without Fermi surface fine-tuning. Third, distinct from vertical stripes constrained by cylinder widths on regular lattices, the diagonal lattice permits infinite-length stripes (i-stripes) on cylinders of appropriate width, circumventing the finite-size limitations of stripes. 
The formation of i-stripes, which is geometrically prohibited on regular lattices, reveals unprecedented physics regarding unconventional SC and charge density orders in Hubbard model. 
Our large-scale DMRG simulations demonstrate a 2D-like PDW emerged in i-stripe phase (Fig.~\ref{Fig:phase_diagram}(c)), providing probably the first controlled numerical evidence of dominant PDW in the single-band Hubbard model. 
Moreover, the resultant quantum phase diagram containing the evolution of unconventional SC and charge density orders offers new microscopic insights into competing orders.


{\bf Model and Method: }%
We employ large-scale DMRG \cite{White1992, Schollwock2005} calculation with GPU-acceleration to investigate ground-state properties of the hole-doped Hubbard model on diagonal square lattice defined as%
\begin{eqnarray}\label{Eq:Ham}
H =- \sum_{ij\sigma} t_{ij} \left(\hat{c}^\dagger_{i\sigma} \hat{c}_{j\sigma} + h.c.\right)+ U\sum_i \hat{n}_{i\uparrow}\hat{n}_{i\downarrow}, 
\end{eqnarray}
where $\hat{c}^\dagger_{i\sigma}$ ($\hat{c}_{i\sigma}$) represents the electron creation (annihilation) operator on site $i$ with spin $\sigma$, $\hat{n}_{i\sigma}=\hat{c}^\dagger_{i\sigma}\hat{c}_{i\sigma}$ and $\hat{n}_i=\sum_{\sigma}n_{i\sigma}$ are the electron number operators. The electron hopping amplitude $t_{ij}=t$ if $i$ and $j$ are NN and $t_{ij}=t^\prime$ if $i$ and $j$ are NNN. $U$ denotes the on-site Coulomb repulsion. We set $t=1$ as the energy unit and lattice spacing $a=1$. 
The $\pi/4$-rotated lattice is shown in Fig.\ref{Fig:phase_diagram}(a), with open boundary condition along $\hat{e}_1=(\sqrt{2},0)$ and periodic condition along $\hat{e}_2=(0,\sqrt{2})$. The unit vectors of the canonical square lattice are denoted as $\hat{e}_x=(1/\sqrt{2},-1/\sqrt{2})$ and $\hat{e}_y=(1/\sqrt{2},1/\sqrt{2})$. In addition to translation symmetries, the diagonal lattice also respects the mirror symmetry about $\hat{e}_1$.
The lattice contains $N=2 \times L_1\times L_2$ sites with $L_1$ ($L_2$) unit cells along the $\hat{e}_1$ ($\hat{e}_2$) direction. The concentration of doped holes is defined as $\delta=\frac{N_h}{N}$ with $N_h=N-N_e$ representing the number of holes measured from half-filling.

We focus on $L_2=6$ and $8$ cylinders with length up to $L_1=20$ and doping $\delta=5\%$ to $15\%$. In the $U=0$ limit, the Fermi surface (Fig.~\ref{Fig:phase_diagram}(b)) is determined by the single-particle dispersion $\epsilon(k)=-2t(\cos k_x+\cos k_y)-2t^\prime (\cos(k_x+k_y)+\cos(k_x-k_y)) + \mu$, where $\mu$ denotes the chemical potential and $k=(k_x, k_y)$ is in the BZ of the regular square lattice. The dashed lines represent quantized momenta of the diagonal lattice which contains a nodal line of $d$-wave SC. 
For $U=8$ and $12$ cases, we perform over 100 sweeps and keep up to $m=48000$ states in each DMRG block to obtain reliable results with truncation error $\epsilon\lesssim 5\times 10^{-5}$. Further numerical details are provided in the Supplementary Materials (SM).

{\bf Quantum phase diagram: } The hole-doped Hubbard model ($t'<0$) with $U=12$ exhibits an unexpectedly rich quantum phase diagram across $\delta=5\%$ to $15\%$. 
At low doping, the system develops diagonal charge and spin stripes with locked ordering vector $\mathbf{Q}^c_{dia}=2 \mathbf{Q}^s_{dia}=(4\pi \delta, 4\pi \delta)$ depicted in Fig.~\ref{Fig:phase_diagram}(c1), accompanied by a short-range $d$-wave SC.
Increasing doping to $\delta\sim 10\%$ leads to a WC* phase characterized by the bidirectional charge density order with wavevectors $\mathbf{Q}^c_{h,1}=(0, 2\pi/3)$ and $\mathbf{Q}^c_{h,2}=(2\pi/3-\delta Q,\delta Q)$ \cite{White2004,Jiang2024}, where $\delta Q$ quantifies small deviations from ideal crystallization. However, unidirectional spin stripes shown in Fig.~\ref{Fig:phase_diagram}(c2) signifies a separation between spin and charge degrees of freedom, i.e. a hallmark of holon. SC correlations in WC* phase remain short-ranged but develop unexpected sign-alternating oscillations at long distances.

In the $\delta \gtrsim 12\%$ regime, the charge density profile exhibits infinite-length stripes across the entire cylinders (Fig.~\ref{Fig:phase_diagram}(c3)). This long stripe state restores translation symmetry along $\hat{e}_x$ as evidenced by the disappearance of $\mathbf{Q}^c_{h,2}$ peak in Fourier transform of charge density $n(k)$. 
Reduced competition between charge order and SC coherence leads to quasi-long-range SC correlations in both $\hat{e}_x$ and $\hat{e}_y$ directions, implying the SC ordering is beyond 1-D physics. Intriguingly, the oscillatory short-range SC in WC* phase evolves to a fully developed PDW in the i-stripe phase---providing probably the first definitive numerical evidence of dominant PDW in the single-band Hubbard model. 

This doping-induced quantum phase diagram persists in a wide parameter region across $U=8\sim12$ and $t'=-0.1\sim -0.3$ on $L_2=6$ cylinders. Remarkably, the evidences of the i-stripe phase are also observed in wide systems with realistic model parameters. For example, for the cuprate-related parameters $\delta=1/8$ and $t'= -0.2 \sim -0.3$, we identify period-4 ``half-filled'' charge i-stripes with locked period-8 spin stripes on $L_2=8$ cylinders. The associated pair-pair correlation along the i-stripes displays dominate oscillation with wavelength $\sim 4$ (see SM for details), notably aligned with the PDW observed in Bi2212 and Bi2201 systems \cite{Hamidian2016, Ruan2018, Li2021}. These results suggest that the emerged i-stripe$+$PDW phase are stable in the cuprate-related models on wide systems.

{\bf The i-stripe$+$PDW phase:} We investigate the physical properties of the i-stripe phase by calculating the $t'= -0.3$, $U=12$ and $\delta=14.2\%$ model on $L_2=6$ cylinders. As illustrated in the upper panel of Fig.~\ref{Fig:phase_diagram}(c3), the charge density profile of this model exhibits a period-3 modulation that spontaneously breaks the translation symmetry and mirror symmetry about $\hat{e}_1$, assembling two independent i-stripes spanning through cylinders. 
The modified spin density profile $\tilde{S}^z(r_1,r_2) =(-1)^{\sqrt{2}r_1} S^z(r_1,r_2)$ in the lower panel displays an AFM background with anti-phase domain walls across each i-stripes. The local spin along domain walls host density-wave structure with wavelength $\sim 3.5$, distinct from the period-2 AFM observed on primitive lattices. 
The stability of the i-stripe phase is further confirmed by its slightly lower local energy compared to the ``2/3-filled'' stripe state on regular lattices under same model parameters (see SM).

\begin{figure}[bt]
\centering
\includegraphics[width=\linewidth]{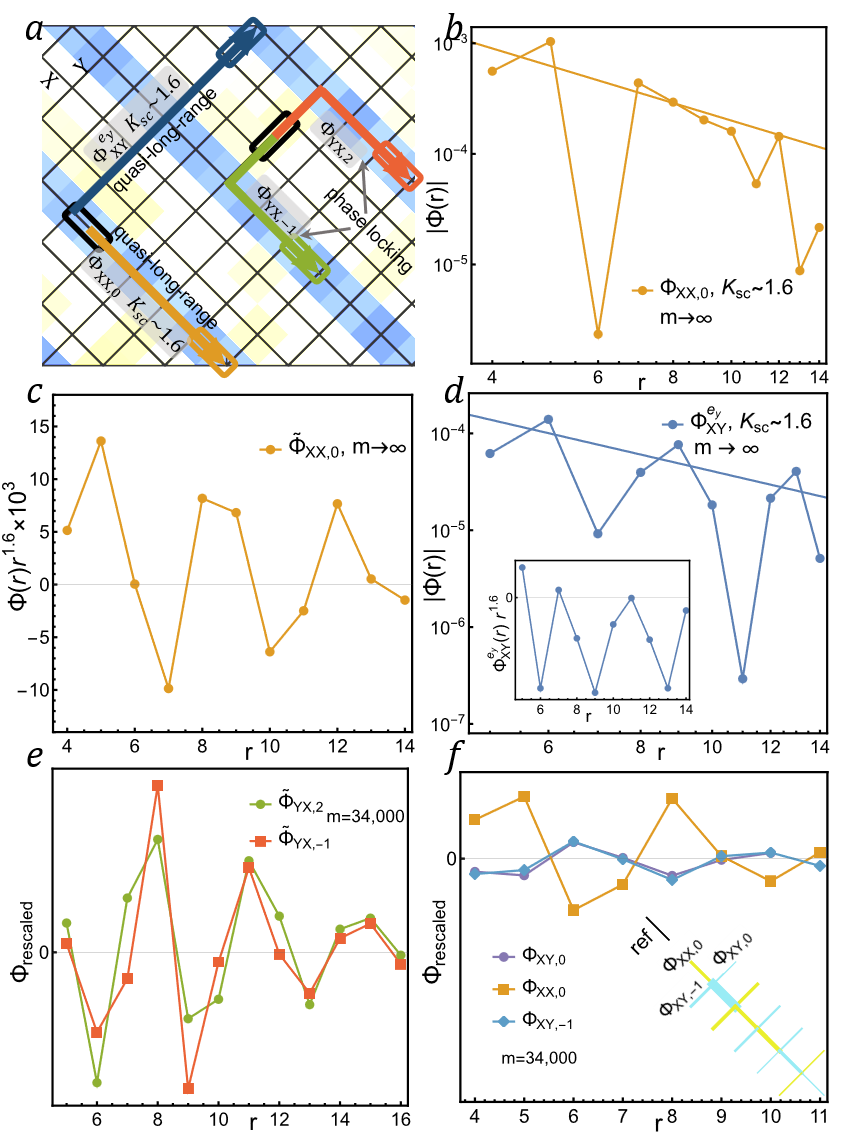}
\caption{(Color online) (a) Sketch of the four typess of SC correlation functions $\Phi_{\text{XX},0}$, $\Phi^{e_y}_{\text{XY}}$, $\Phi_{\text{YX},0}$ and $\Phi_{\text{YX},0}$ measured in (b-e). (b) Correlation $\Phi_{\text{XX},0}$ (orange) along i-stripe, obtained by finite truncation-error scaling to $m\rightarrow \infty$ limit. The envelope is fitted by power-law function with exponent $K_{sc}\sim 1.6$. (c) The rescaled $\tilde{\Phi}_{\text{XX},0}(r) = \Phi_{\text{XX},0}(r)/r^{-K_{sc}}$ exhibiting spatial oscillation with wavelength $\sim 3.5$. (d) The SC correlation ${\Phi}_\text{XY}^{\hat{e}_y}$ perpendicular to the i-stripes (blue), decaying with same exponent $K_{sc}\sim 1.6$. Inset: the rescaled $\tilde{\Phi}_\text{XY}^{\hat{e}_y}(r)$.
(e) Spatial oscillations of $\tilde{\Phi}_{\text{YX},2}$ and $\tilde{\Phi}_{\text{YX},-1}$ on two adjacent stripes. (f) Three correlations ${\Phi}_{\text{XY},0}$, ${\Phi}_{\text{XY},-1}$ and ${\Phi}_{\text{XX},0}$ supporting local $d$-wave symmetry of the PDW state. The results in (e-f) is measured with $m=34000$ block states.}
\label{Fig:pdw}
\end{figure}

The elongated stripes that facilitate charge density fluctuations provide a new arena for exploring the potential unconventional SC order. We analyze the SC properties by following pair-pair correlators shown in Fig.~\ref{Fig:pdw}(a)
\begin{eqnarray}
    \Phi _{\alpha \beta, \delta_y}(x) &=&\left< \Delta _{\alpha,\hat{r}_0}^\dagger \Delta _{\beta,\hat{r}_0 + x \hat{e}_x+\delta_y \hat{e}_y} \right> , \label{Eq:SC1} \\
    \Phi_{\alpha \beta}^{\hat{e}_y}(y)&=&\left< \Delta_{\alpha,\hat{r}_0}^\dagger \Delta_{\beta,\hat{r}_0 + y \hat{e}_y} \right> , \label{Eq:SC2}
\end{eqnarray}
where $\Phi _{\alpha \beta, \delta_y}$ and $\Phi_{\alpha \beta}^{\hat{e}_y}$ describe SC correlations along and perpendicular to i-stripes, respectively. Here, the i-stripes are set along $\hat{e}_x$ for simplicity.
$\hat{\Delta} _{\alpha, \hat{i}}^{\dagger} =\frac{1}{\sqrt{2}}\left( c_{\uparrow ,\hat{i} }^\dagger c_{\downarrow ,\hat{i} +\hat{e}_\alpha}^\dagger-c_{\downarrow ,\hat{i}}^\dagger c_{\uparrow ,\hat{i} + \hat{e}_\alpha}^\dagger \right)$ creates spin-singlet pairs on the $\alpha=$ X or Y bond originating from site $\hat{i}$. $\delta_y$ in Eq.~\ref{Eq:SC1} denotes the displacement along $\hat{e}_y$.
Our central finding, based on various types of SC correlations depicted in Fig.~\ref{Fig:pdw}, is that a $d$-wave PDW with divergent susceptibility establishes in the i-stripe phase. 


Spatial analysis of SC correlations $\Phi_{\text{XX},0}$ along the i-stripe (orange line in Fig.~\ref{Fig:pdw}(a)) reveals a pronounced quasi-long-range PDW that can be captured by the function $\Phi_{\text{XX},0}(x)=f(x)\tilde{\Phi}_{\text{XX},0}(x)$. The envelope $f(x) \sim x^{-K_{sc}}$ exhibits power-law decaying with exponent $K_{sc}\sim 1.6$ as shown in Fig.~\ref{Fig:pdw}(b), yielding divergent SC susceptibility $\chi_{sc} \sim T^{- (2 - K_{sc}) }$ as temperature $T\rightarrow 0$. The oscillatory component $\tilde{\Phi}_{\text{XX},0}(x)$ represented in Fig.~\ref{Fig:pdw}(c) can be well described by $\cos(Q_{p} x+ \theta)$ with negligible uniform SC contribution. The ordering vector $Q_{p}$ is roughly $0.55 \pi$, corresponding to a PDW wavelength $\sim 3.5$. 
The i-stripes and oscillatory SC correlations persist on wider cylinders with $L_2=8$. 
For instance, we observe robust ``half-filled'' i-stripes with period-4 PDW in cuprate-relevant $1/8$-doped model with $t'=-0.2\sim -0.3$ and $U=12$ (see SM for details).

For SC pairs separated along $\hat{e}_y$ (blue line in Fig.~\ref{Fig:pdw}(a)), their correlation ${\Phi}_\text{XY}^{\hat{e}_y}(y)$ is also quasi-long-range. As shown in Fig.~\ref{Fig:pdw}(d), the envelop of ${\Phi}_\text{XY}^{\hat{e}_y}(y)$ exhibits similar power-law decaying $y^{-K^{\hat{e}_y}_{sc}}$ with exponent $K^{\hat{e}_y}_{sc}\sim K_{sc}$. The same SC exponents of correlations along both $\hat{e}_x$ and $\hat{e}_y$ implies that the SC order is beyond 1D physics and might persist when system approaches to 2D limit. We note that perpendicular SC correlations are also measurable on wide regular lattices but exhibits only short-range SC correlations according to previous DMRG studies of hole-doped ($t'<0$) Hubbard models \cite{Jiang2021White,Jiang2022White,Jiang2024,Lu2023Sign}. This suggests that surpassing the constraint of stripes length is crucial to enhance superconductivity in Hubbard model. 

The spatial distribution of ${\Phi}_\text{XY}^{\hat{e}_y}(y)$ is illustrated in the inset of Fig.~\ref{Fig:pdw}(d). Instead of the sign-changing oscillation, we only find a modulation of amplitude of ${\Phi}_\text{XY}^{\hat{e}_y}$ at long distance \cite{Dolfi2015}, which can be attributed to the coupling between charge density stripes and 2D-like PDW. 
Combined ${\Phi}_{\text{XX},0}(x)$ and ${\Phi}_\text{XY}^{\hat{e}_y}(y)$ together, the leading term of SC order parameter can be written as  
\begin{eqnarray}\label{Eq:order}
    \Delta_{SC}(x,y)\sim\cos{(Q_{p}x+\theta)}[\Delta_0+\delta\Delta\cos(Q' y+\theta')] ,
\end{eqnarray}
where $Q'$ describes the modulation along $\hat{e}_y$. $\theta$ and $\theta'$ denote phases of the $\hat{e}_x$ oscillation and $\hat{e}_y$ modulation, respectively. $\delta\Delta < \Delta_0$ represents the spatial modulation induced by the presence of charge stripes. Eq.~\ref{Eq:order} suggest that the phases of PDW oscillations along different i-stripes are identical. To verify this relation, we calculate SC correlations $\Phi_{\text{YX},2}$ and $\Phi_{\text{YX},-1}$ which have same reference bond but opposite $\hat{e}_y$ displacements. The overlap of oscillatory components of two correlations depicted in Fig.~\ref{Fig:pdw}(e) directly confirms the form of order parameter in Eq.~\ref{Eq:order}.


The pair symmetry of the PDW state is diagnosed through comparative analysis of correlations ${\Phi}_{\text{XX},0}$, ${\Phi}_{\text{XY},0}$ and ${\Phi}_{\text{XY},-1}$ illustrated in Fig.~\ref{Fig:pdw}(f). By comparing amplitudes of the three correlations, we find that the relationship ${\Phi}_{\text{XY},0}(r) \sim {\Phi}_{\text{XY},-1}(r) \sim -\alpha {\Phi}_{\text{XX},0}(r)$ (with $\alpha < 1$) approximately hold for all distance $r$, confirming the local $d$-wave pair symmetry.

\begin{figure}[bt]
\centering
\includegraphics[width=\linewidth]{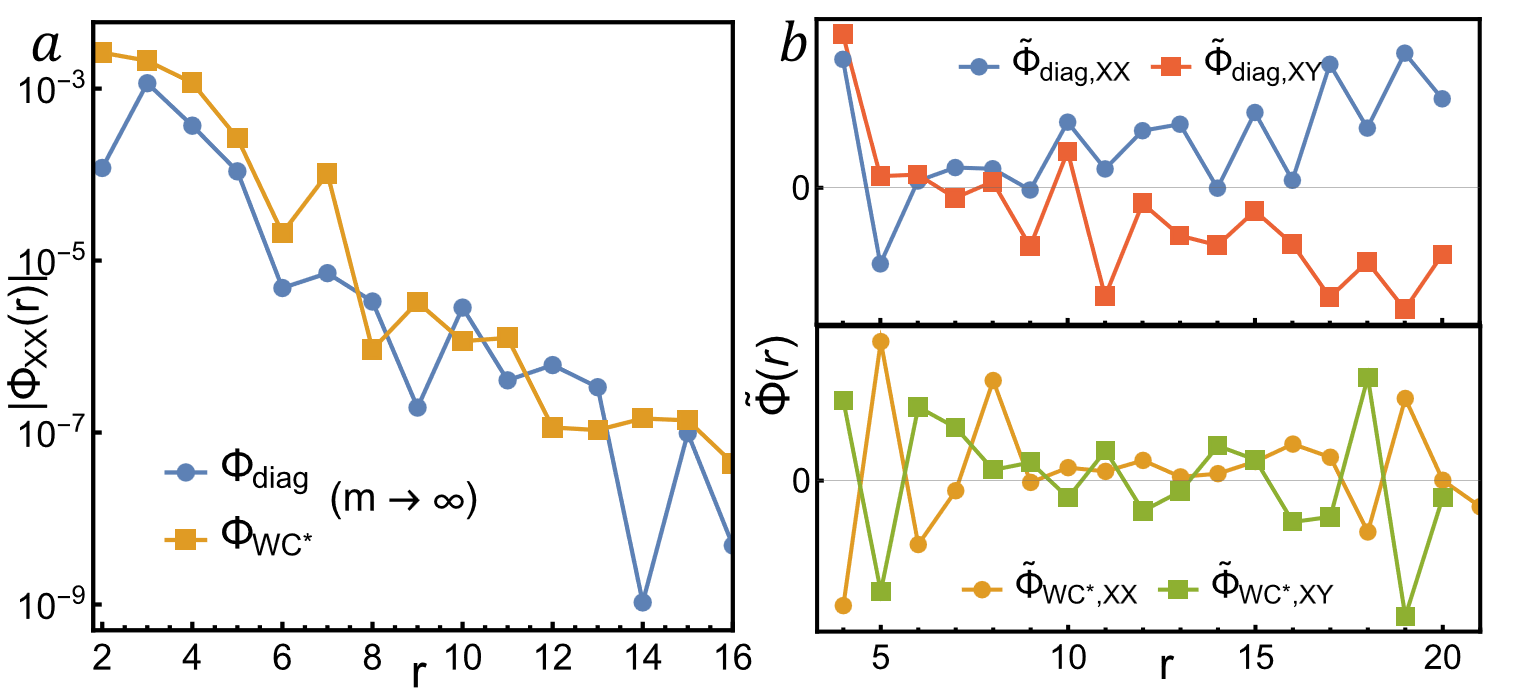}
\caption{(Color online) (a) Short-range pair correlations $\Phi_{\text{XX},0}$ in $t'=-0.3$ and $U=12$ models with $\delta=6.25\%$ (diagonal stripe phase) and $\delta=10\%$ (WC* phase). The correlation length $\xi_{sc} \sim 1.5$. (b) The rescaled pair correlation $\tilde{\Phi}(r) \sim e^{0.7r} \Phi(r)$ in diagonal stripe and WC* phases. }
\label{Fig:SC_all}
\end{figure}

{\bf Evolution of stripe and SC orders: } After determined properties of the i-stripe$+$PDW phase, we systematically explore the evolution of charge density order and SC correlations as the hole doping increases. 
At low doping $\delta \lesssim 9\%$, the system forms diagonal charge density stripes which preserve the translation symmetry along $\hat{e}_2$ and mirror symmetry about $\hat{e}_1$.
The corresponding spin stripes also exhibits anti-phase domain walls across density stripes which leads to the locked relation $2 \mathbf{Q}^s_{dia} = \mathbf{Q}^c_{dia}=(4\pi \delta, 4\pi \delta)$ shown in Fig.~\ref{Fig:phase_diagram}(c1). Note that these diagonal stripes have length of $\sqrt{2}L_2$, much shorter than i-stripes. 

Increasing the doping level to $\delta \sim 10\%$ induces a bidirectional CDW order manifested in arrays of hole-rich (blue) spots assembling an enlarged square lattice. As shown in Fig.~\ref{Fig:phase_diagram}(c2), the enlarged lattice exhibits small deviation from the ideal configuration, which can be quantified by $n(k)$ peaks at $\mathbf{Q}^c_{h,1}=\{0, 2\pi/3\}$ and $\mathbf{Q}^c_{h,2}=\{2\pi/3-\delta Q, \delta Q\}$, with small $\delta Q\sim 0.1\pi$ for $\delta=10\%$. In contrast, the spin density remains unidirectional stripes (lower panel), demonstrating spin-charge separation consistent with previous observation of WC* phase in $t'<0$ Hubbard models on regular square lattice \cite{Jiang2024}. Our result confirms the robustness of WC* phase in the hole-doped regime of the phase diagram. Intriguingly, the $\hat{e}_x$ component of $\mathbf{Q}_{h,2}^c$, $2\pi/3-0.1\pi=0.56\pi$, is nearly identical to the $Q_p$ of PDW order in Eq.~\ref{Eq:order}, implying an intimate relation between charge density fluctuation and PDW ordering.

\begin{figure}[bt]
\centering
\includegraphics[width=\linewidth]{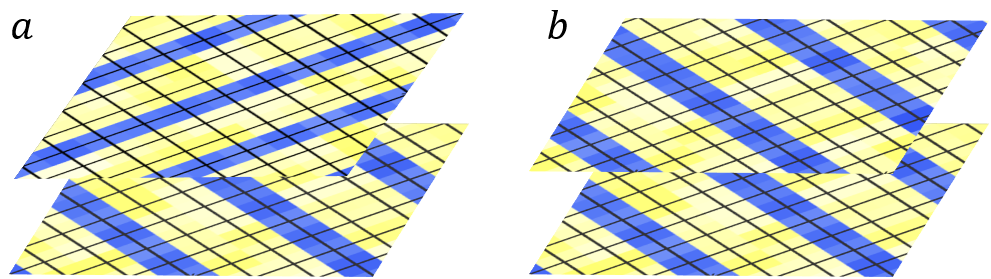}
\caption{(Color online) (a) Perpendicular and (b) parallel staking of two charge-stripe planes.}
\label{Fig:layer}
\end{figure}

To further elucidate this relationship, we examine the evolution of pair correlation $\Phi_{\text{XX},0}$ from diagonal stripe to i-stripe phase.
Fig.~\ref{Fig:SC_all}(a) shows that both diagonal stripe and WC* phases host short-range pair correlations, supporting that fully developed short stripes and Wigner crystal order significantly suppress SC ordering. However, the rescaled correlations $\tilde{\Phi}_{\text{XX},0}$ and $\tilde{\Phi}_{\text{XY},0}$ in Fig.~\ref{Fig:SC_all}(b) indicate that sign-changing oscillation of SC has already developed in the WC* phase. These oscillations are further intensified during the transition from WC* to the i-stripe phase, and eventually leads to dominant PDW along i-stripes. In contrast, in the diagonal stripe phase, we only find a uniform $d$-wave SC satisfied the relation $\tilde{\Phi}_{\text{XX},0}>0>\tilde{\Phi}_{\text{XY},0}$ at long distance. 

{\bf Discussion: } 
Our large-scale DMRG simulations reveal a rich quantum phase diagram for the hole-doped Hubbard model: at $\delta \lesssim 9\%$, a diagonal stripe phase with short range SC; near $\delta \sim 10\%$, a WC* phase with bidirectional charge order; and above $\delta \gtrsim 12\%$, an infinite-length stripe phase hosting dominant 2D-like PDW order. 
This PDW order, characterized by quasi-long-range SC correlations in both lattice directions and sign-changing oscillation along stripe direction, provides probably the first controlled numerical evidence of PDW physics in single-band Hubbard models. 

PDW superconductivity is widely recognized as a key factor in explaining layer decoupling in stripe-ordered superconductors like LBCO \cite{Li2007, Berg2007} and SrTa$_2$S$_5$ \cite{Devarakonda2024}, where vanishing interlayer Josephson coupling is attributed to geometric frustration arising from the PDW order on neighboring charge-stripe planes. Typically, within each layer, the SC order parameter is uniform along stripes but changed sign between adjacent stripes \cite{Berg2007}. In stark contrast, our study reveals distinct PDW pattern with intrinsic spatial oscillations along each i-stripe. Considering the PDW order described in Eq.~\ref{Eq:order}, the leading term of Josephson coupling between $j$ and $j+1$ layers is given by
\begin{eqnarray}
    C= c_0 \int d^2r(\Delta^*_j\Delta_{j+1}+h.c.) ,
\end{eqnarray}
where $\Delta_j = \Delta_{SC}(x,y,\theta_j,\theta_{j}')$ and $c_0$ is a constant. For orthogonally stacked layers (Fig.~\ref{Fig:layer}(a)), the SC order transforms as $\Delta_{j+1} = \Delta_{SC}(y,-x,\theta_{j+1},-\theta_{j+1}')$ on $j+1$ layer, which induces a destructive interference that yields $C=0$, providing an alternative view of imaging the suppression of interlayer coupling in cuprates. For parallel stacking (Fig.~\ref{Fig:layer}(b)), the integral simplifies to $C \sim \cos(\theta_{j}-\theta_{j+1}) [2+\cos(\theta'_{j}-\theta'_{j+1})]$, which value vanishes when PDW phase difference satisfies $\theta_{j}-\theta_{j+1}=\pi/2$.

On regular square lattice, the limited stripe length ($\leq L_y$) caused by the geometric constraint fundamentally restricts measurement of long-distance correlations along the stripe direction. Consequently, prior DMRG studies could only investigate correlations perpendicular to stripes, which exhibited short-range SC correlations without sign-changing oscillations in hole-doped models on cylinders with moderate width \cite{Jiang2021White,Jiang2022White,Jiang2024,Lu2023Sign}. 
Our diagonal lattice design overcomes this limitation through the formation of extended i-stripes. The direct probe of long-distance physics along i-stripe reveals quasi-long-range PDW which is unobserved in the previous studies. The extended stripe formation also significantly strengthens the SC correlation perpendicular to stripes, establishing a 2D-like SC order with quasi-long-range SC correlations in both directions.
Moreover, it uncovers a novel interplay between CDW and PDW order: the short-range sign-changing SC correlation in the WC* phase evolves into quasi-long-range PDW upon partially melting of density order, suggesting that fluctuating charge density orders along the stripe could be crucial in facilitating PDW, even though fully developed CDW order suppresses SC in WC* phase. 
Similar implications regarding density-fluctuation enhanced superconductivity have been proposed in several studies for uniform SC \cite{Castellani1995,Perali1996,Arrigoni2004}. Nonetheless, a comprehensive understanding of the multifaceted connection between the spin/charge density fluctuation and the enhancement of finite-momentum SC pairing demands further investigations. 
Finally, by spontaneously breaking the reflection symmetry of diagonal square lattices, it is possible to realize the nematic phase which can not be precisely determined on regular square cylinders. All these new observations establish the $t'$-Hubbard model on diagonal square lattices as a novel platform for exploring PDW physics and the interplay between charge density, spin density, and unconventional SC in high-Tc superconductors and related quantum materials.

\emph{Acknowledgments:} We would like to thank Steven A. Kivelson, Hong-Chen Jiang, Zheng-Cheng Gu and Hong Yao for insightful discussions and suggestions. Y.-F.J. acknowledges support from National Key R$\&$D Program of China under Grants No. 2022YFA1402703 and from NSFC under Grant No. 12347107 and 12574160. The DMRG code used in this work is publicly available in Github \cite{code}

\bibliography{refs}

\clearpage

\begin{widetext}

\renewcommand{\theequation}{S\arabic{equation}}
\setcounter{equation}{0}
\renewcommand{\thefigure}{S\arabic{figure}}
\setcounter{figure}{0}
\renewcommand{\thetable}{S\arabic{table}}
\setcounter{table}{0}

\section{Supplemental Material}

\subsection{Numerical details}
Most numerical results presented in this work are obtained through DMRG simulations employing up to $m=48000$ block states with over 100 sweeps, leveraging GPU acceleration for computationally intensive matrix operations ubiquitous in DMRG simulations. To reduce truncation errors in the finite-bond-dimension calculation, we perform the finite-truncation-error extrapolation for representative physical quantities using data from $m=34000$ to $48000$ block states. 
In Fig.\ref{AFig:trunc_error}, we show the detailed extrapolation applied to ground-state energy $E_{GS}$, single particle correlation functions $G(r)$, and pair-pair correlation functions $\Phi(r)$ for the $14\%$ doped models with $t'=-0.3$ on $L_2=6$ cylinder. 
The extrapolation procedure employs a second-order polynomial 
\begin{equation}
    O(\epsilon) = O_0 + a_1 \epsilon + a_2 \epsilon^2 ,
\end{equation}
where $O$ represents the physical quantities, $\epsilon$ denotes the truncation error associated with the number of states $m$, and $a_1$, $a_2$ are fitting parameters. Fig.\ref{AFig:trunc_error}(a) illustrates the fitting procedure for extracting the ground-state energy $E_{GS}(\epsilon \rightarrow 0)$ from a series of energy measured at finite $m$. For the spatial correlation functions $G(r)$ and $\Phi(r)$, second-order polynomial extrapolations are independently applied to the data at each $r$ to extract correlation functions in the zero truncation-error limit, as shown in Fig.\ref{AFig:trunc_error}(b) and (c).

\begin{figure}[bth]
\centering
\includegraphics[width=0.8\linewidth]{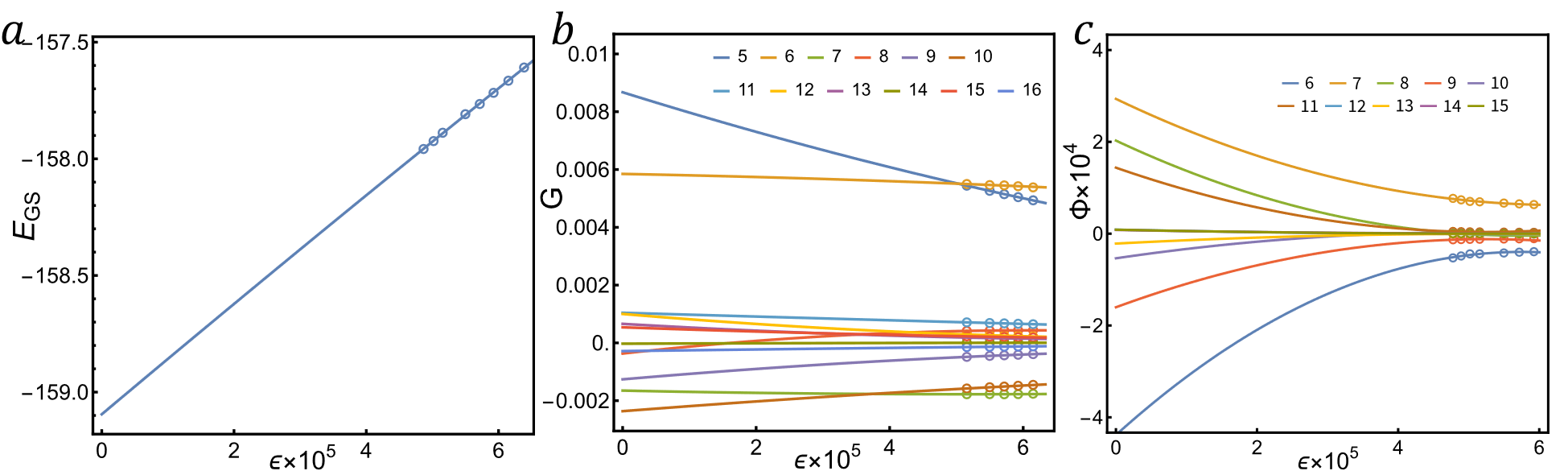}
\caption{(Color online) The finite-truncation-error extrapolation for (a) ground-state energy $E_{GS}$, (b) single particle correlation functions $G(r)$, and (c) pair-pair correlation functions $\Phi_{\text{XX},0}(r)$ for the $14\%$ doped model with $t'=-0.3$ on $L_2$ cylinder. The data points are obtained from DMRG simulations with $m$ up to $48000$ block states.}
\label{AFig:trunc_error}
\end{figure}

\subsection{Local energy of model on diagonal and regular lattices}
To confirm the PDW state observed in the main text as an competitive ground-state candidate, we conduct a comparative calculation of local energies for $t'=-0.3$, $U=12$ model on diagonal and regular square lattices. The system size of the diagonal (regular) lattice is $2 L_1=40$ $(L_x=40)$ and $L_2=6$ $(L_y=6)$. Previous studies of the $t'=-0.3$ Hubbard model on the $L_y=6$ regular lattice established charge density stripes with wavelength $\lambda_{c}=2/(3\delta)$ across a range of doping concentrations. To minimize the influence from open boundaries, we consider commensurate stripes with wavelength $\lambda_c=5$ by setting doping concentration at $\delta=13.33\%$, and calculate the average local energy through the expectation value of the local Hamiltonian in 15-site supercells ($3\times5$ sites) illustrated in Fig.~\ref{AFig:localenergy}(a). To improve the accuracy, we further perform averaging over a set of $3\times5$ clusters generated by translation along the $\hat{e}_y$ direction. This yields a local energy density $E_{reg}=-0.6599$ per site.

For the diagonal square lattice. Though the doping concentration $\delta = 13.33\%$ is slightly deviate from the $\delta \sim 14\%$ case studied in the main text, the infinity-length stripes persist, as shown in Fig.~\ref{AFig:localenergy}(b). Since the wavelength of the stripes is 3 lattice spacing, we calculate the energy of the local Hamiltonian in a $7\times 3$ subsystem illustrated in the right part of Fig.~\ref{AFig:localenergy}(b). Following the same procedure, we find the average local energy is $E_{dia}=-0.6612$ per site on diagonal lattice, slightly lower than $E_{reg}$ for the regular lattice. Both of the energies are measured with $m=30000$ DMRG block states.

\begin{figure}[bth]
\centering
\includegraphics[width=0.75\linewidth]{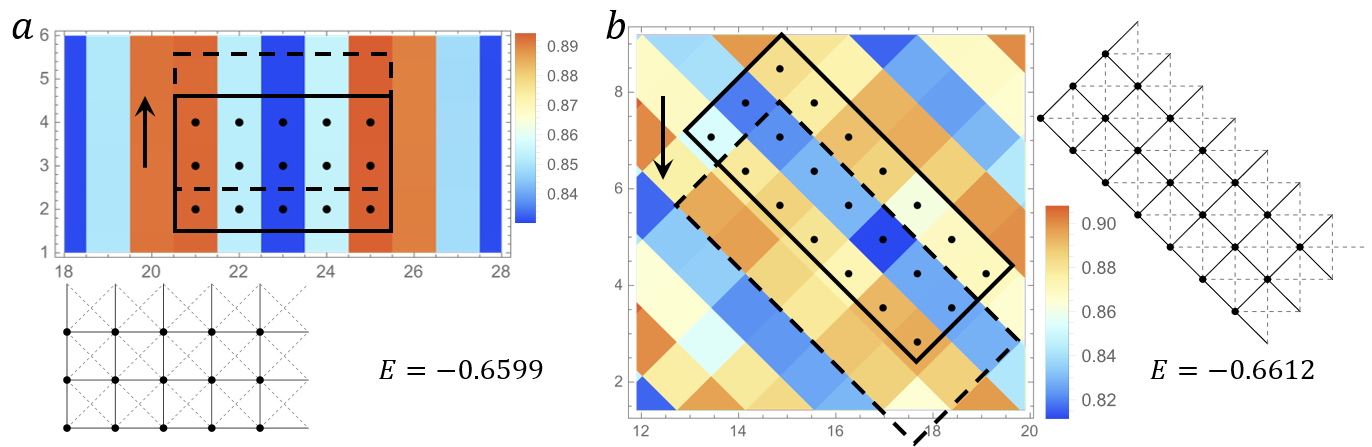}
\caption{(Color online) (a) Local energy of the $t'=-0.3$, $\delta=13.33\%$ and $U=12$ model on regular square lattice with $L_x=40$ and $L_y=6$. The upper part shows the charge density profile in the bulk, exhibiting a clear period-5 stripe. The lower part is the illustration of the cluster used to calculate the local energy. Black dots represent the sites, solid and dashed bonds denote the NN and NNN hopping included in the calculation. (b) Local energy of the same model on diagonal square lattice calculated in the same procedure.}
\label{AFig:localenergy}
\end{figure}

\subsection{Results for $U=8$ models}

\begin{figure}[bth]
\centering
\includegraphics[width=0.9\linewidth]{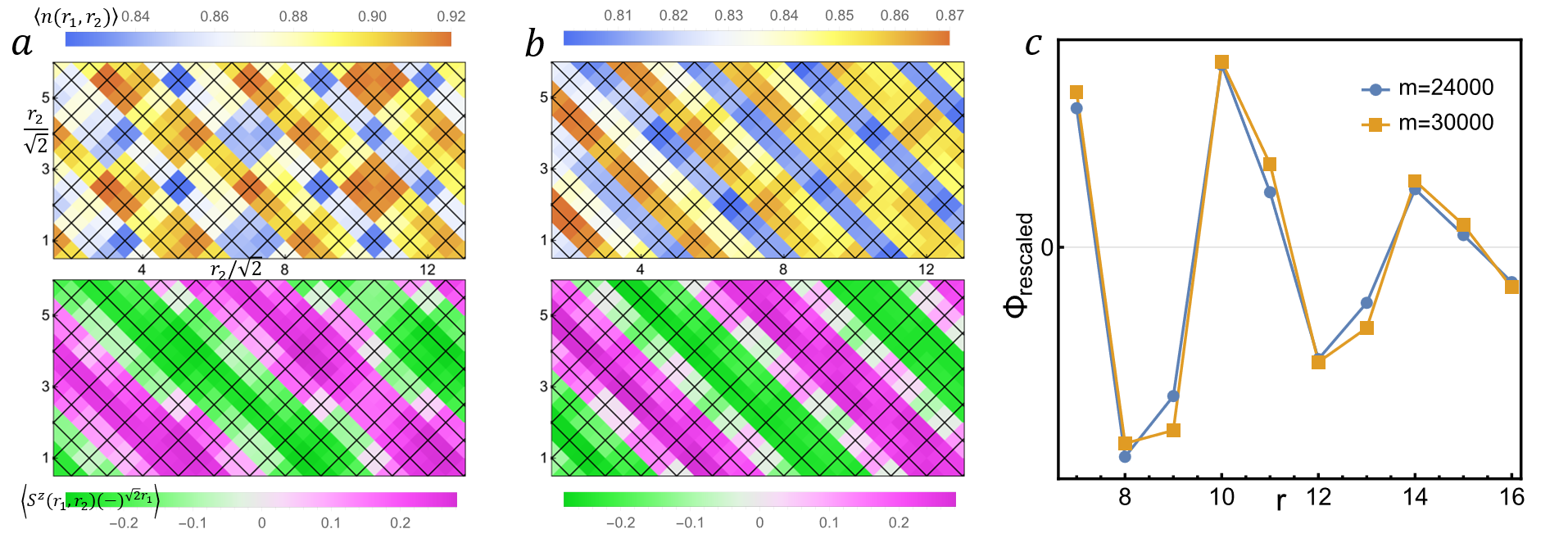}
\caption{(Color online) The ground-state properties of the $t'=-0.3$ and $U=8$ Hubbard model on diagonal square lattice with $L_2=6$: (a) Charge and spin density profiles of the $\delta=10\%$ model. The patterns are consistent with WC* state. (b) Charge and spin density profiles of the $\delta=15.8\%$ model, confirming the i-stripe phase in $U=8$ model. (c) The rescaled pair-pair correlation functions measured with $m=24000$ and $30000$ states in the $\delta=15.8\%$ and $U=8$ model.}
\label{AFig:U=8}
\end{figure}

We investigate the ground-state properties of the $t'=-0.3$ and $U=8$ Hubbard model on $L_y=6$ diagonal square lattice. 
As shown in Fig.~\ref{AFig:U=8}(a), at $\delta=10\%$ doping, the charge and spin density profiles are consistent with the WC* state discussed in the main text. The lattice constant of the crystal of holons is around 3. The i-stripe phase is also observed in the $U=8$ model. Fig.~\ref{AFig:U=8}(b) shows the charge and spin density profiles obtained for $\delta=15.8\%$ model, where we can find infinite-length density and spin stripes spanning the entire lattice. The wavelength of the stripes is $\sim 3$ lattice spacing, identical to those observed in the $U=12$ Hubbard model. 
Further analysis of the pair-pair correlation function measured along the stripe reveals a clear sign-changing oscillation. 
To compensate for rapid decay of correlation functions measured with limited number of states ($m\le30000$), we rescaled the correlations using power law functions to highlight their oscillation in Fig.~\ref{AFig:U=8}(c). Crucially, increasing the bond dimension from $m=24000$ to $30000$ leads a clear amplification of the oscillation at long distance.
These results indicates that the main physical results discussed in the main text persist as Hubbard repulsion reduces from $U=12$ to $8$, the main change of the phase diagram for $U=8$ models is that the infinite-stripe phase is shifted to a slightly higher range of doping concentration.

\subsection{Results on $L_2=8$ cylinders with $m=40000$ block states}
   
To rule out the possible finite size effect in the i-stripe$+$PDW phase, we investigate the ground-state properties of the Hubbard model on wider $L_2=8$ cylinders. Due to the significantly increased entanglement entropy on wider cylinders, applying finite-truncation-error extrapolation for the $L_2=8$ systems becomes extremely challenging. Therefore, we restrict our discussion on the physical properties measured with large bond dimension up to $m=40000$. Fig.~\ref{AFig:8-leg} shows the low-energy properties of $U=12$ and $\delta=12.5\%$ model on diagonal lattice with $L_2=8$, $L_1=15$. For both $t'=-0.2$ and $-0.3$ cases, we can see that the charge density converges to the period-4 ``half-filled'' stripe, and the spin density pattern forms AFM stripes with anti-phase domain walls. The wavelength of spin stripes is $8$ lattice spacing, which is twice that of charge stripes. 

\begin{figure}[bth]
\centering
\includegraphics[width=0.8\linewidth]{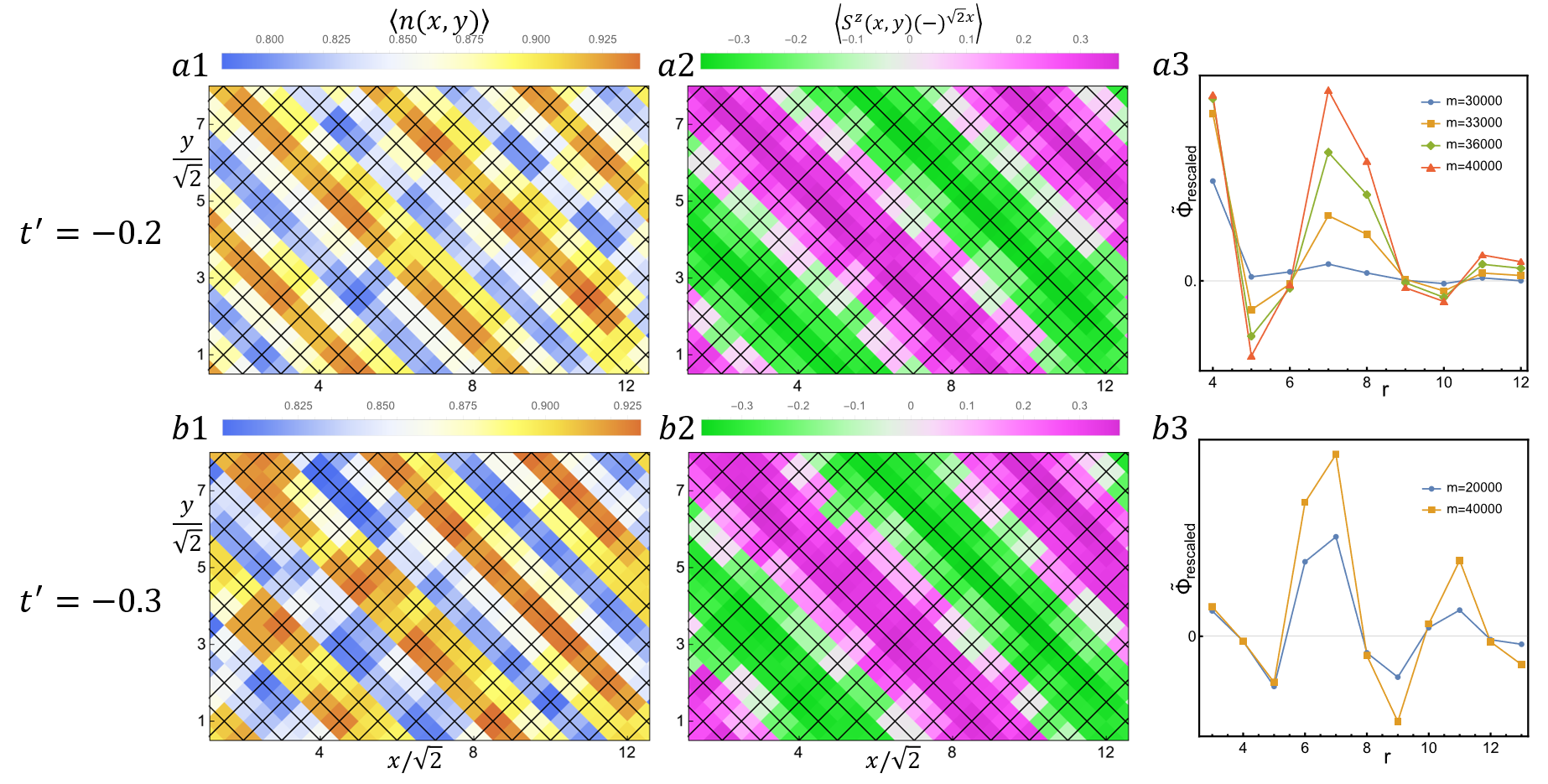}
\caption{(Color online) (a1-a2) Charge density and spin profiles of $\delta=12.5\%$ doped model with $t'=-0.2$ on $L_2=8$ cylinders. The DMRG block dimension is $m=40000$. (a3) Rescaled pair-pair correlation function $\Phi_{\text{YY},0}$ measured along the density stripes shown in (a1). (b1-b3) Charge density profile, spin density profile and pair-pair correlation functions for the $t'=-0.3$ and $\delta=12.5\%$ model on $L_2=8$ cylinders.}
\label{AFig:8-leg}
\end{figure}

We further calculate the pair-pair correlation function along i-stripes demonstrated in Fig.~\ref{AFig:8-leg}(a1) and (b1). As the width of the stripes increases on $L_2=8$ cylinders, we are now able to measure the SC correlation $\Phi_{\text{YY},0}$ between two Y bonds. 
Similar to the $U=8$ case, we rescaled the correlations using same power-law function to emphasize their spatial oscillation with wavelength $\sim 4$ lattice spacing. It can be clearly found in Fig.~\ref{AFig:8-leg}(a3) and (b3) that as the bond dimension increases from $m=20000$ to $40000$, the amplitude of the PDW oscillation is significantly amplified, implying that the PDW could persist on wider cylinders.

\subsection{Charge and spin density order along the stripe}
Here, we present the residual spin $S_z$ along an infinite-length stripe in the stripe phase. We choose the same parameter as the representative point of stripe phase mentioned in the main text, namely the $14\%$ doped Hubbard model with $t'=-0.3$ on $L_2=6$ cylinder. In Fig.~\ref{AFig:SDW_CDW} (a), we replot the spin density profile $\left< S^z(r_1,r_2) \right>$ without removing the AFM background to illustrate weak SDW along the anti-phase domain walls of the AFM background (black line). As shown in Fig.~\ref{AFig:SDW_CDW} (b), along the black line, the residual spin $S_z$ exhibits a weak density-wave pattern with a wavelength $\sim 3.5$ lattice spacing, which is distinct from the simple period-2 AFM pattern observed in the Hubbard model on a regular square lattice.

\begin{figure}[bth]
\centering
\includegraphics[width=0.95\linewidth]{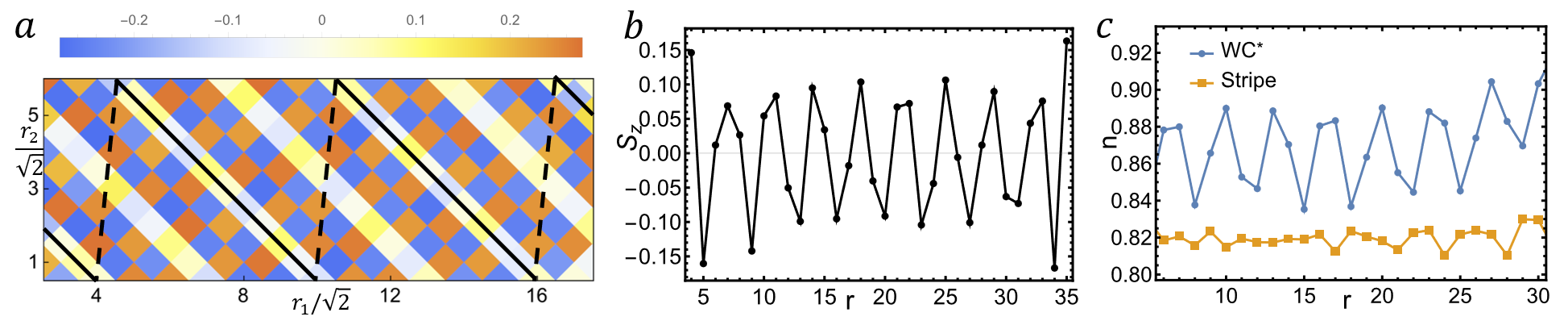}
\caption{(Color online) (a) The spin profile $\left < S^z(x,y)\right >$ of the $14\%$ doped Hubbard model with $t'=-0.3$ and $L_2=6$. Note here we removed the factor $(-1)^{\sqrt{2}x}$ in the spin profile. Black line represents one of the anti-phase domain walls. (b) The weak spin density wave along the domain wall. (c) The charge density $\left < n(r)\right >$ along the black line in the model with doping $\delta=10\%$ and $14\%$, which shows the establishment of charge density wave on the stripe as the system evolves from WC* phase to stripe phase.}
\label{AFig:SDW_CDW}
\end{figure}

As discussed in the main text, during transition from WC* to i-stripe phase, the density modulation associated with momentum $\mathbf{Q}_{h,2}^c$ vanishes. To depict this evolution, we compare charge densities along the $e_{x}$ direction (black line in Fig.~\ref{AFig:SDW_CDW} (a)) for both $\delta=10\%$ doping (WC* phase) and $\delta=14\%$ doping (i-stripe phase) on $L_2=6$ cylinder. As shown in Fig.~\ref{AFig:SDW_CDW} (c), in contrast to the nearly uniform density distribution in the i-stripe phase, the charge density in WC* phase exhibit clear spatial oscillation with period $\sim 3.5$ along the same line. Surprisingly, direct comparison between Fig.~\ref{AFig:SDW_CDW} (b) and (c) reveals that, the SDW ordering/fluctuation in i-stripe phase and the CDW ordering in WC* phase share almost identical spatial distributions.

\subsection{Single-particle correlation function in the i-stripe phase}

We measure the single-particle correlation $ G_{\sigma \sigma'}(r) = \left< c_{i,\sigma}^\dagger c_{i+r,\sigma'} \right>$ in the i-stripe$+$PDW phase. In Fig.~\ref{AFig:singleparticle}, we show the correlation $G_{\uparrow \uparrow}(r)$ measured along the stripes in the $t'=-0.3$ and $\delta=14\%$ model with up to $m=42000$ DMRG block states. At long distances, its decaying behavior can be fitted by a exponential function $G(r)\sim e^{-r/\xi_G}$ with correlation length $\xi_G \sim 3.3$ lattice spacing (see Fig.~\ref{AFig:singleparticle}(a)), or fitted by a power-law function $G(r)\sim r^{-K_G}$ with exponent $K_G \sim 2.6$ (see Fig.~\ref{AFig:singleparticle}(b)).

\begin{figure}[bth]
\centering
\includegraphics[width=0.7\linewidth]{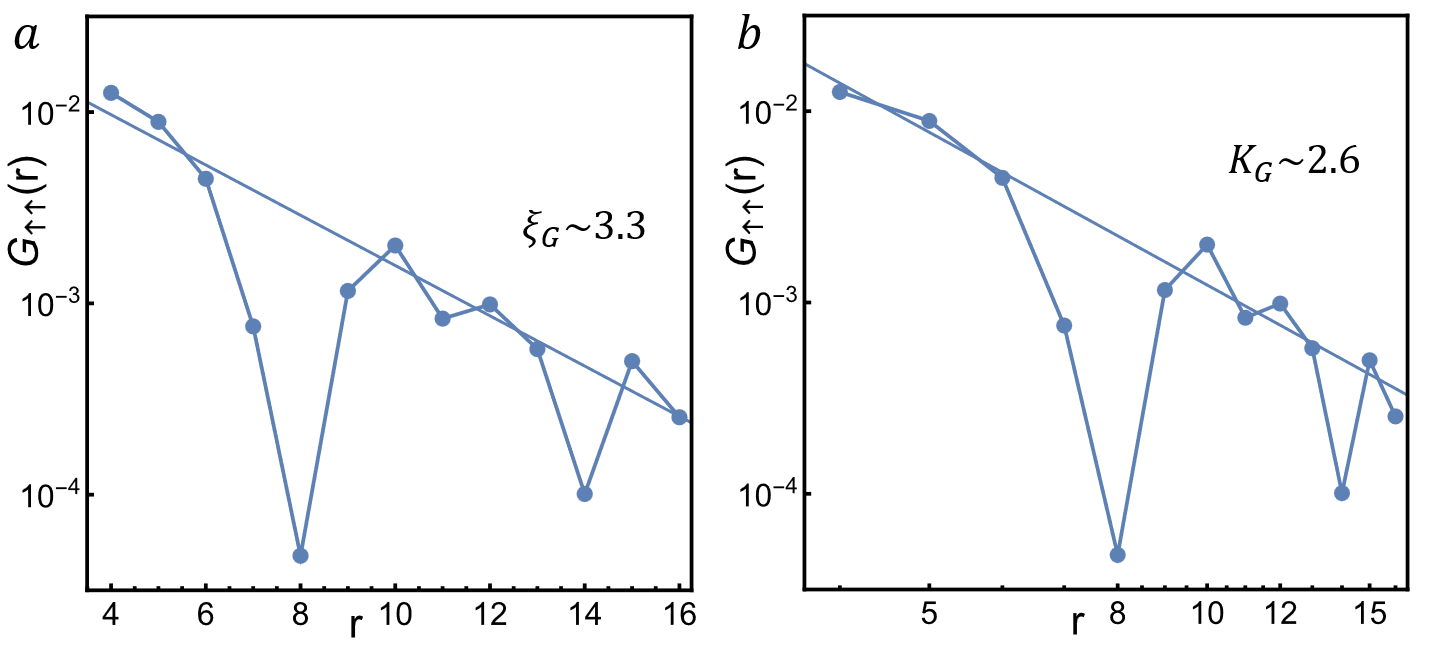}
\caption{(Color online) Single particle correlation function of $\delta=14\%$ doped models with $t'=-0.3$ on $L_2=6$ cylinders.}
\label{AFig:singleparticle}
\end{figure}

\subsection{Stable stripe phase in a range of $t'$}
In the main text, our study focuses on the model with $t'=-0.3$. Here, we demonstrate in Fig.~\ref{AFig:stripe_tprime} that the i-stripe phase remains stable over a wide range of $t'$ from $-0.1$ to $-0.3$. In the cases of $t'=-0.1$ and $-0.2$, we observe spin and charge stripes exhibiting similar wavelength and characteristics as those in the $t'=-0.3$ model.
However, for larger $t'=-0.4$ case, while the charge density profile remains in stripe patterns, the spin stripes break into many smaller domains, suggesting potentially new bidirectional phases in the large negative $t'$ region. Conversely, for $t'=0$ case (the pure Hubbard model), no evidences of stripe orders is observed even with $m$ up to 36000 DMRG block states; hence it will be intriguing to investigate the properties of converged ground-state of the $t'=0$ Hubbard model on wider diagonal square lattice in future work.

\begin{figure}[bth]
\centering
\includegraphics[width=0.8\linewidth]{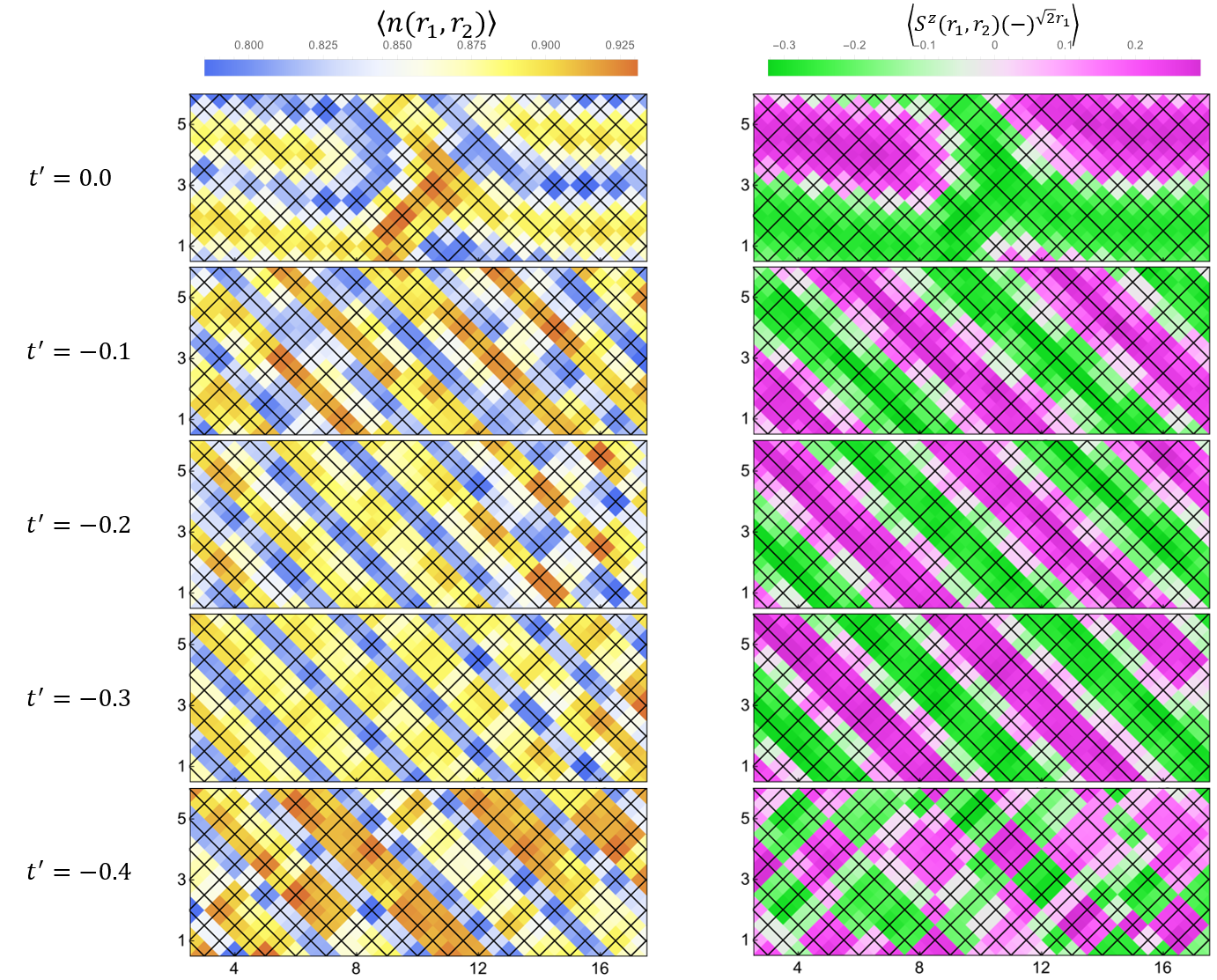}
\caption{(Color online) Charge density and spin profiles of $U=12$ and $\delta=14\%$ doped models with a series of $t'=0.0$ to $-0.4$ on $L_2=6$ cylinders.  }
\label{AFig:stripe_tprime}
\end{figure}

\end{widetext}

\end{document}